\documentclass[a4paper,twoside,12pt,english]{article}

\usepackage[T1]{fontenc}
\usepackage[latin1]{inputenc}
\usepackage{babel}
\usepackage{amsmath}
\usepackage{latexsym}
\usepackage{graphicx}

\def\beq{\begin{equation}}
\def\eeq{\end{equation}}
\def\bitem{\begin{itemize}}
\def\eitem{\end{itemize}}
\def\bear{\begin{array}}
\def\ear{\end{array}}

\def\d{\partial}

\def\m{\hat{m}}

\def\2f{{\phi}^2}
\def\ds5{ds_{(5)}}

\begin{document}


\begin{titlepage} \vspace{0.2in} 

\begin{center} {\LARGE \bf 
Dimensional  Reduction of the 5D Kaluza-Klein Geodesic Deviation Equation}\\ 
\vspace*{1cm}
{\bf  V. Lacquaniti $^{1,2,3,\diamond}$, G. Montani $^{1,4,5,\dag}$, F. Vietri $^{1,2,\star}$ }\\
\vspace*{1cm}
$^1$ ICRA---International Center for Relativistic Astrophysics, 
Physics Department (G9),
University  of Rome, "`La Sapienza"', 
Piazzale Aldo Moro 5, 00185 Rome, Italy.\\

$^2$  Physics Department  "`E.Amaldi "`,  University of Rome, "`Roma Tre"', 
Via della Vasca Navale 84, I-00146, Rome , Italy \\ 

$^3$ LAPTH -9, Chemin de Bellevue BP 110 74941 Annecy Le Vieux Cedex, France \\

$^4$ ENEA- C. R. Frascati ( Department F. P. N. ), via E.Fermi 45, I-00044, Frascati, Rome, Italy \\
$^5$ ICRANET - C.C. Pescara, Piazzale della Repubblica 10, I-65100, Pescara, Italy \\

\vspace*{1.8cm}

{\bf   Abstract  \\ } \end{center} \indent
In the work of Kerner et al. (2001) the problem of the geodesic deviation in a 5D Kaluza Klein background is faced. The 4D space-time projection of the resulting equation coincides with the usual geodesic deviation equation in the presence of the Lorenz force, provided that the fifth component of the deviation vector satisfies an extra constraint which takes into account the $q/m$ conservation along the path. The analysis was performed setting as a constant the scalar field which appears in Kaluza-Klein model. Here we focus on the extension of such a work to the model where the presence of the scalar field is considered. Our result coincides with that of Kerner et al. when the minimal case $\phi=1$ is considered, while it shows some departures in the general case. The novelty due to the presence of $\phi$ is that the variation of the $q/m$ between the two geodesic lines is not conserved during the motion; an exact law for such a behaviour has been derived.

\vfill
\hrule

$^{\diamond}$ valentino.lacquaniti@icra.it,
$^{\dag}$ montani@icra.it,
$^{\star} $ vietrif@fis.uniroma3.it

\end{titlepage}

\section*{Introduction}
Kaluza-Klein (KK) theories (\cite{kaluza}-\cite{overduinwesson}) offer in vacuum a  tool to consider the unification of interactions in a geometrical picture; 
in the 5D compactified model  we reproduce the coupling of the gravitational field with a U(1) vector field, which we identify with the electromagnetic field, plus an extra scalar field, which is the scale factor of the extra dimension (\cite{jordan}-\cite{cianfrani}).  The most striking results of such a model are the explanation for the generation of the local U(1) gauge symmetry and the discretization of the charge (\cite{marroccomontani}, \cite{cianfrani}, \cite{KK}).
Although these results are encouraging, nevertheless  the introduction of matter in the compactified model  is still an open problem: it yields some inconsistencies in the dynamics, like the generation of a huge massive modes spectrum and the  problem concerning the  proper identification of the correct coupling factors associated to gauge currents (\cite{overduinwesson}, \cite{KK}, \cite{vandongen}). 
In order to add a contribute  to this debate, in this paper we consider the dimensional reduction of the geodesic deviation equation in the framework of the compactified 5D KK model. Such a topic was faced at first by Kerner et al. (\cite{kerner}), setting from the beginning as a constant the scalar field, and the following equation has been derived:

\beq 
\frac{D^2 \delta x^\alpha }{Ds^2} = -R^{\alpha}_{{}{}\beta\gamma\lambda} u^{\beta} \delta x^{\gamma} u^{\lambda} +\frac{ek w_5}{\sqrt{1+w^2_5}} \delta x^{\nu} \nabla_{\nu} \left(F^{\alpha\beta} u_{\beta}\right)-ekF^{\alpha}_{\nu} u^{\nu}\delta Q
\label{geodev1} \, .
\eeq
Here $ek$ is a dimensional constant which reads $(ek)^2=\frac{4G}{c^2}$, $w_5$ is the fifth component of the 5D velocity  $w^{A}=\frac{dx^A}{ds_5}$, $u^\rho$ is the 4D velocity, and as usual $R^{\alpha}_{{}{}\beta\gamma\lambda}$ and $F^{\mu\nu}$ are the Riemann and the Faraday tensor respectively.
The factor $\frac{ek w_5}{\sqrt{1+w^2_5}}$ defines the ratio $q/m$ of a given particle and the above equation  describes the behaviour of the deviation parameter $\delta x^{\mu}$ for two close test particles with ratios $q/m$ and $q/m+\delta Q$.
The factor $\delta Q$ explicitly reads $
 \delta Q=\frac{d \delta x_5 }{ds} - F_{\mu\nu} \delta x^{\mu} u^{\nu}
\label{quinta componte  phi = 1 }
$, and its dynamics is governed by the condition:

\beq 
\frac{d}{ds}\left(\frac{d \delta x_5 }{ds} - F_{\mu\nu} \delta x^{\mu} u^{\nu}\right)=0\,.
\label{51}
\eeq
Therefore, choosing as suitable initial condition the value  $\delta Q= 0$   the following equation can be written:
\beq 
\frac{D^2 \delta x^{\alpha} }{Ds^2} = -R^{\alpha}_{\beta\gamma\lambda} u^{\beta} \delta x^{\gamma} u^{\lambda} + \frac{q}{m}\delta x^{\nu} \nabla_{\nu} \left(F^{\alpha\beta} u_{\beta}\right).
\label{rif}
\eeq
In the usual 4D theory it describes the behaviour of the deviation parameter $\delta x^{\alpha}$ between two close test-particles, with same charge $q$ and mass $m$, in presence of an electromagnetic field.
Setting as a constant the scalar field however,  does not allow us to distinguish between a $4$D theory from a  $5$D theory; then the topic of this work is to build a $5$D  deviation equation in presence of the scalar field, in order to achieve a relation which could be in principle  used   to search evidence of an extra dimension.

\section{Geodesic equation in Kaluza Klein}

Let us consider at first the reduction of the geodesic equation in the framework of the cylindrical and compactified KK model.
 We  assume, as  in the 4D theory, that the motion of the free particle follows a geodesic line, \textit{i.e.} the dynamics is governed by the following  Action:
 
\beq
S=-\m\int ds_5.
\label{malemale}
\eeq
Here $\m$ is an unknown mass parameter and  $\ds5$ reads
\beq
ds_5^2=ds^2-\phi^2\left(ekA_{\mu}dx^{\mu}+dx^5 \right)^2\, , \quad\quad \mu=0,1,2,3,
\eeq 
where $ds$ is the 4D line element and $\phi$ is the extra scalar field.
The reduction of the $5$D geodesic equation $\frac{dw^{A}}{ds_5}=0$ provides the following set:

\beq 
\frac{Du^{\gamma}}{Ds}=ek\frac{w_5}{\sqrt{1+\frac{w^2_5}{\phi2}}}F^{\gamma\beta} u_{\beta} +\frac{1}{\phi^3}\left( u^{\gamma} u^{\beta} - g^{\gamma\beta}\right )\d_{\beta} \phi\left( \frac{w^2_5}{1+\frac{w^2_5}{\phi^2}}\right )
\label{geodkkred}
\eeq

\beq
\frac{dw_5}{ds}=0
\label{u5}
\eeq
Equation (\ref{geodkkred}) describes the motion of the reduced particle, which is now interacting with the electromagnetic field and the scalar one. Equation (\ref{u5}) sets the existence of a constant of  motion;  it is possible also to show that $w_5$ is a scalar object.
Explicitly we have:
\beq
w_5=-\phi^2\left( ekA_{\mu}w^{\mu}+w^{5}   \right)
\label{w5}
\eeq
Such a constant allows us to define the electromagnetic coupling. Indeed, we define the charge-mass ratio as follows:
\beq  
ek\frac{w_5}{\sqrt{1+\frac{w^2_5}{\phi2}}} = \frac{q}{mc^2}.
\label{qm}
\eeq
 An insight concerning the role of $\m$ can be achieved by considering the hamiltonian formulation and  studying  the associated dispersion relation; while the study of the motion equation allows us to define only the charge-mass ratio, the analysis of conjugate momenta gives us a definition for mass and charge separately ( see \cite{KK} for details  ). It turns out that the interacting 4D particle is characterized by the following relation:

\beq
\left(P_\mu\ - qA_\mu \right)\left(P^\mu\ - qA^\mu \right)\ = m^2\, ,
\eeq
where:
\beq
q=ekP_5 \, ,\quad\quad m^2=\m^2+\frac{P_5^2}{\phi^2}\, .
\eeq
In these formulas $P_{A}=\m w_{A}$ are  the conjugate momenta and these definitions reproduce the $q/m$ ratio previously considered . But, more important, we see that the mass parameter $\m$ we put in the Action (\ref{malemale}) does not correctly  represent the physical mass of the particle which is indeed $m$.
The multiplying factor $\alpha=\sqrt{1+\frac{w_5^2}{\phi^2}}$ that enters into the definition of the mass arises because we are forced, during the procedure, to perform a reparemetrization  between $ds$ and $ds_5$, such that we have :
\beq 
\alpha=\sqrt{1+\frac{w^2_5}{\phi^2}} =\frac{ds}{ds_5}\, .
\label{riparametrizzazione ds5 a ds4}
\eeq
 We will see, in the next section, how this factor plays a role in the dynamical equation governing of the geodesic deviation.
\section{Geodesic deviation equation in Kaluza-Klein }
The starting point of our analysis is the 5D geodesic deviation equation, which we obtain extending to the 5D background the usual geodesic deviation equation; indeed, it is possible to show that, from the analysis of two close geodesic lines, the following equation can be built, 

\beq 
\frac{D^2 \delta x^A }{Ds_5^2} = -^5R^A_{{}{}BCD} w^B \delta x^C w^D\, ,
\label{eqgen}
\eeq
where $^5R^A_{{}{}BCD}$ is the 5D Riemann tensor and $\delta x^A$ represent the $5$D displacement factor between two geodesic equations.
Via the dimensional reduction procedure the 4D part of equation (\ref{eqgen}) yields the formula:

$$ 
\frac{D^2 \delta x^\alpha }{Ds^2} \ -\left(\frac{w^2_5}{\ 1+\frac{w^2_5}{\phi^2}}\right)\frac{1}{{\phi}^3}\frac{d\phi}{ds}\frac{D\delta{x}^\alpha}{ds}\ = \ - R^\alpha_{\beta\gamma\lambda} u^\beta \delta x^\gamma u^\lambda \ +
$$
$$   
\ + {\frac{w_5}{\sqrt{1\ +\frac{w^2_5}{{\phi^2}}}}}\delta x^\nu \nabla_\nu \left[F^{\alpha\beta} u_\beta \right] \ +
\left(\frac{w^2_5}{\ 1+\frac{w^2_5}{\phi^2}}\right)\delta x^\nu \nabla_\nu \left(\frac{\partial^\alpha\phi}{\phi^3}\right)\ - 
$$
\begin{equation}
\ -\left(F^\alpha_\nu u^\nu \ - 2\frac{w_5}{\sqrt{1\ + \frac{w^2_5}{\phi^2}}}\frac{ \partial^\alpha \phi}{{\phi}^3}\right)\delta Q\, .
\label{boh}
\end{equation} 
In the same way the reduction of the fifth component yields:
\beq
\frac{d}{ds}\left[\sqrt{1\ +\frac{w^2_5}{\phi^2}} \left(\delta Q
\right)\right] \ = - w_5\frac{\delta x^\nu \partial_\nu \phi}{\phi^3}\frac{d \phi}{ds}\, .
\label{quinta componente caso generale}
\eeq
The definition of $\delta Q$ is now as follows:
\beq
\delta Q=\left({\frac{d \delta x_5 }{ds}} \ - 2\frac{\delta x_5}{\phi}\frac{d \phi }{ds} - {\phi^2} F_{\mu\nu} \delta x^\mu u^\nu \ + 2\frac{w_5}{\sqrt{1\ +\frac{w^2_5}{\phi^2}}} \frac{\delta x^\rho \partial_ \rho\phi}{\phi}\right)\, .
\label{orka2}
\eeq
Now we are going to demonstrate that $\delta Q$ represent the variation of the  charge-mass ratio in presence of a scalar field. From the definition of $\frac{q}{m}$ in (\ref{qm}) we have:

\beq
\delta \left(\frac{q}{m}\right) \approx ek\frac{\delta w_5}{\sqrt{1+\frac{w^2_5}{\2f}}} + O\left(w^2_5\right).
\label{orka1}
\eeq
Let us now calculate $\delta w_5$ by the definition (\ref{w5}):

$$
\delta w_5 \ = \delta \left[\2f\left(ekA_\mu u^\mu \ +  w^5\right)\right]\ = \delta \left(\2f\right)\left[ekA_\mu u^\mu \ +  w^5\right]\ + \2f\delta \left(ekA_\mu u^\mu \ +  w^5\right)\ = 
$$

\beq
\ = 2w_5 \frac{\delta x^\rho \partial_\rho \phi}{\phi}\ +ek\2f\left(\delta x^\rho \partial_\rho A_\mu u^\mu \ -\frac{dA_\mu \delta x^\mu }{\ds5}\right)\ +\frac{d\left(ek\2fA_\mu \delta x^\mu\ +\2f\delta x^5 \right)}{\ds5}\ - \frac{d\2f}{\ds5}\left(ekA_\mu \delta x^\mu \ +\delta x^5 \right).
\label{legamephi}
\eeq
The second term can be rewritten as the contraction of the Faraday tensor with the quantities $\delta x^\rho$ and $u^\mu$, while the last two terms represent the  fifth component of the vector $\delta x_A$. Then we have:  

\beq 
\delta w_5 =
{\frac{d \delta x_5 }{\ds5}} \ - 2\frac{\delta x_5}{\phi}\frac{d \phi }{\ds5} - {\phi^2} F_{\mu\nu} \delta x^\mu w^\nu \ +2w_5\frac{\delta x^\rho \partial_ \rho\phi}{\phi}.
\label{pinco2}
\eeq
The last step is to use the factor $\alpha=\sqrt{1+\frac{w_5^2}{\phi^2}}$  to perform a reparemeterization  between $ds$ and $ds_5$:
$$
\delta w_5 =
\sqrt{1\ +\frac{w^2_5}{\2f}}\left[{\frac{d \delta x_5 }{ds}} \ - 2\frac{\delta x_5}{\phi}\frac{d \phi }{ds} - ek{\2f} F_{\mu\nu} \delta x^\mu u^\nu \ +\frac{2w_5}{\sqrt{1\ +\frac{w^2_5}{\2f}}}\frac{\delta x^\rho \partial_ \rho\phi}{\phi}\right].
$$ 
Using the definition (\ref{orka1}) we finally  find the (\ref{orka2}):

\beq
 \delta \left(\frac{q}{m}\right)\ = \delta Q\ = \left[{\frac{d \delta x_5 }{ds}} \ - 2\frac{\delta x_5}{\phi}\frac{d \phi }{ds} - ek{\2f} F_{\mu\nu} \delta x^\mu u^\nu \ +\frac{2w_5}{\sqrt{1\ +\frac{w^2_5}{\2f}}}\frac{\delta x^\rho \partial_ \rho\phi}{\phi}\right].
\label{pinco1}
\eeq
Therefore, eq. (\ref{boh}) describes the geodesic deviation equation in presence of electromagnetism plus scalar field, while eq. (\ref{quinta componente caso generale}) describes the behaviour of $\delta Q$ or the influence of the scalar field on the ratio charge to mass ratio along the path of motion. 
Recalling the definition $\alpha=\sqrt{1+\frac{w^2_5}{\phi^2}} =\frac{ds}{ds_5}$ we can manipulate eq. (\ref{quinta componente caso generale}) and  write:

\beq
\frac{d\alpha}{ds} \ = \ -\alpha\frac{u^2_4}{\phi^3}\frac{d \phi}{ds} 
\eeq

\beq
\delta{\alpha} \ = \alpha\left(\frac{1}{\phi^2}\frac{d\delta x_4}{ds}-\frac{u_4}{\phi^3}\delta x^\nu \partial_{\nu}\phi \right).	
\eeq
The combination of this two terms allows to analyze the eq. (\ref{quinta componente caso generale}) from another point of view and to rewrite it as follows:

\beq
\frac{d}{ds}\delta Q=\left[   \frac{\phi}{w_5^2}\frac{d}{ds}\delta x_5-\alpha\delta\alpha\frac{\phi^3}{w_5^2}-\frac{\delta Q}{\alpha}       \right]\frac{d}{ds}\alpha.
\label{www}
\eeq
The above equation thus gives the deviation of the factor $\delta Q$  in term of the reparametrization factor $\alpha$; in such a  way we can link the problem of the not conservation of the charge-mass ratio with the projection factor from $\ds5$ to $ds$. 

\section{Concluding Remarks}

As we can see from equation (\ref{www}), the factor $\delta Q$ is no more conserved during the motion, due to the variation of $\alpha$ induced by the variation of $\phi$. Therefore, even though we assume  initial condition $\delta Q=0$, nevertheless it does not vanish during the motion and it affects  the path deviation, as we can see via eq. (\ref{boh}). It is remarkable that such a variation can be addressed to the variation of $\alpha$, which is indeed the same factor that causes the ill-defined mass  that affects the compactified model inducing the generation of  the Planckian massive tower (\cite{overduinwesson},\cite{KK},\cite{vandongen}).
This is the main non-trivial result we get, which generalises the findings of the paper (\cite{kerner}).
Furthermore, from a phenomenological point of view, it provides a testable tool for an experimental detection of the field $\phi$, through an analysis of its tidal effects on particle dynamics. It is worth noting, in this respect, that the variation of $\delta Q$, being related to $\alpha$, is connected to the variation of the relative size between the extra dimension and the ordinary 4D lengths. This outcome is a direct consequence of the fact that we adopted a geodesic approach to describe the motion of the unreduced 5D test particle in our compactified model.
Other approaches to the test particles dynamics in the framework of extra dimensional physics   adopt different schemes concerning the properties of the extra dimension. Usually such approaches  relax the cylindrical and the compactification hypotheses, 
and deal with non compact space and embedding procedure (\cite{rub1}-\cite{shap2}) -which at the end leads to the brane scheme (\cite{rand1},\cite{rand2})-, or still adopt the compactification hypothesis but face the problem of the motion using more than 5D and invoking a symmetry breaking mechanism for the generation of mass (\cite{collins}). In a recent proposal (\cite{KK},\cite{KK2}),  it has been suggested that the geodesic approach is not the proper one when dealing with test particle, due to the violation of the 5D Principle of Equivalence; hence, it was proposed to face the particle motion via the conservation law of a 5D cylindrical matter tensor defining the particle via an appropriate Papapetrou expansion (\cite{pap}). With respect to this debate the issue of the geodesic deviation can  give an interesting theoretical tool for a comparison of the models. At the same time, the extension to multidimensional non-abelian model appears an interesting issue to be pursued.

\section*{Acknowledgements} 
Authors are grateful to CGM members for their feedback.
The work of V. Lacquaniti has been partially supported by a fellowship "Bando Vinci" granted from the "French-Italian University".

\end{document}